\documentclass[%
 aip,
 jmp,%
 amsmath,amssymb,
 reprint,%
]{revtex4-1}

\usepackage{graphicx}
\usepackage{dcolumn}
\usepackage{bm}

\begin{document}

\preprint{AIP/123-QED}

\title[Internal disruptions and sawtooth like activity in LHD]{Internal disruptions and sawtooth like activity in LHD}

\author{J. Varela}
 \email{jvrodrig@fis.uc3m.es}
\affiliation{Universidad Carlos III, 28911 Legan\'es, Madrid, Spain}
\author{L. Garcia}
\affiliation{Universidad Carlos III, 28911 Legan\'es, Madrid, Spain}
\author{S. Ohdachi}
\affiliation{National Institute for Fusion Science, Oroshi-cho 322-6, Toki 509-5292, Japan}
\author{K.Y. Watanabe}
\affiliation{National Institute for Fusion Science, Oroshi-cho 322-6, Toki 509-5292, Japan}
\author{R. Sanchez}
\affiliation{Universidad Carlos III, 28911 Legan\'es, Madrid, Spain}

\date{\today}

\begin{abstract}
LHD inward-shifted configurations are unstable to resistive MHD pressure-gradient-driven modes. These modes drive sawtooth like events during LHD operation. In this work, we simulate sawtooth like activity and  internal disruptions in order to improve the understanding of these relaxation events and their effect over the device efficiency to confine the plasma, with the aim to improve the LHD present and future operation scenarios minimizing or avoiding the disadvantageous MHD soft and hard limits. By solving a set of reduced non-linear resistive MHD equations, we have studied the evolution of perturbations to equilibria obtained before and after a sawtooth like event in LHD. The equilibrium $\beta$  value is gradually increased during the simulation until it reaches the experimental value. Sawtooth like events and internal disruption events take place in the simulation for $\beta_{0}$ values between $1\%$ and $1.48\%$. The main driver of the sawtooth like events is the resonant and non-resonant effect of the $(n=1, m=3)$ mode. The instability is stronger for resonant events, and they only appear when $\beta_{0} = 1.48 \%$. Internal disruptions are mainly driven by the  $(n=1, m=2)$ mode, and  they extend throughout the whole plasma core. Internal disruption events do not show up when resonant sawtooth like events are triggered.
\end{abstract}

\maketitle

\section{Introduction \label{sec:introduction}}

In this work, we investigate magnetohydrodynamic (MHD) mode activity in the Large Helical Device (LHD). The highest stored energies and beta values in LHD are obtained in the so-called inward-shifted configurations.
However, inward-shifted configurations MHD properties are unfavourable because a magnetic hill is located near the magnetic axis, and pressure-gradient-driven modes \cite{1,2} are not stabilized by the magnetic well or the magnetic shear \cite{3}. Previous linear MHD stability studies pointed out that low $n$ modes are unstable \cite{4,5}. In inward-shifted LHD configurations pressure-gradient-driven modes limit the operation efficiency of LHD, but they only increase slightly the energy transport out of the system \cite{7}. There is an interpretation of the stabilizing mechanism that avoids the excitation of low $n$ interchange modes for $ \beta_{0} < 1 \% $; the pressure profile is flattened around the rational surfaces \cite{8,6} where the mode growth saturates. Pressure evolves to a staircase-like profile and the modes suffer periodic excitations and relaxations.

Periodic relaxations events, similar to sawtooth phenomena, have been observed in LHD inward-shifted configurations  \cite{9,10}. This type of activity was detected in pellet fuelled plasmas with peaked pressure profiles and intense NBI heating \cite{11} with and without large net toroidal current \cite{12}. Internal disruptions and sawtooth like activity were first studied in other Stellarator devices like Heliotron-E \cite{13,14,15} and CHS \cite{16,17}. In these devices, sawtooth like events were driven before or after internal disruptions. Internal disruptions strongly affected the equilibrium properties of the system while sawtooth like events could be driven without large changes of the equilibrium characteristics.  In LHD there is only sawtooth like activity experimental evidence, but in future operations with high beta and plasma density in inward-shifted configurations, internal disruptions could be driven and it is important to foresee their effects on the plasma co
 nfinement and their relation with the $1/2$ sawtooth like activity \cite{9}.

The aim of the present research is to simulate internal disruptions and sawtooth like activity to improve the understanding of these relaxation events and try to avoid or minimize their effects over the LHD plasma confinement efficiency. The present and future LHD operation scenarios must be optimized versus the MHD soft and hard limits, because it was observed during the experiments that the MHD activity limits the LHD performance. Advanced LHD operation can be only reached if sawtooth like activity is minimized and if the conditions for internal disruptions are avoided. The present study conclusions show the conditions and consequences of internal disruptions and sawtooth like events over the LHD performance and how to avoid or reduce their effect.

The simulations were made using the FAR3D code \cite{18,21,22}. This code solves the reduced non-linear resistive MHD equations to follow the system evolution under the effect of a perturbation of the equilibrium. The equilibria were calculated with the VMEC code \cite{19}, and they correspond to two LHD configurations without net toroidal current before and after a sawtooth like event  \cite{9}.

This paper is organized as follows. The model equations and the numerical scheme are explained in section \ref{sec:model}. The equilibrium properties, numerical calculation conditions and diagnostics are also described. The simulation results are presented in section \ref{sec:simulation}.  Finally, the conclusions of this paper are presented in section \ref{sec:conclusions}.

\section{Equations and numerical scheme \label{sec:model}}

For high-aspect ratio configurations with moderate $\beta$-values (of the order of the inverse aspect ratio), we can apply the method employed in Ref. 22 for the derivation of the reduced set of equations without averaging in the toroidal angle. In this way, we get a reduced set of equations using the exact three-dimensional equilibrium. In this formulation, we can include linear helical couplings between mode components, which were not included in the formulation developed in Ref. 22.

The main assumptions for the derivation of the set of reduced equations are high aspect ratio, medium $\beta$ (of the order of the inverse aspect ratio $\varepsilon=a/R_0$), small variation of the fields, and small resistivity. With these assumptions, we can write the velocity and perturbation of the magnetic field as
\begin{equation}
 \mathbf{v} = \sqrt{g} R_0 \nabla \zeta \times \nabla \Phi, \quad\quad\quad  \mathbf{B} = R_0 \nabla \zeta \times \nabla \psi,
\end{equation}
where $\zeta$ is the toroidal angle, $\Phi$ is a stream function proportional to the electrostatic potential, and $\psi$ is the perturbation of the poloidal flux.

The equations, in dimensionless form, are
\begin{equation}
\frac{{\partial \psi }}{{\partial t}} = \nabla _\parallel  \Phi  + \frac{\eta}{S} J_\zeta
\end{equation}
\begin{eqnarray} 
\frac{{\partial U}}{{\partial t}} = - {\mathbf{v}} \cdot \nabla U + \frac{{\beta _0 }}{{2\varepsilon ^2 }}\left( {\frac{1}{\rho }\frac{{\partial \sqrt g }}{{\partial \theta }}\frac{{\partial p}}{{\partial \rho }} - \frac{{\partial \sqrt g }}{{\partial \rho }}\frac{1}{\rho }\frac{{\partial p}}{{\partial \theta }}} \right) \nonumber\\
 + \nabla _\parallel  J^\zeta  + \mu \nabla _ \bot ^2U
\end{eqnarray} 
\begin{equation}
\label{peq}
\frac{{\partial p}}{{\partial t}} =  - {\mathbf{v}} \cdot \nabla p + D \nabla _ \bot ^2p + Q
\end{equation}
Here, $U =  \sqrt g \left[{ \nabla  \times \left( {\rho _m \sqrt g {\bf{v}}} \right) }\right]^\zeta$, where $\rho_m$ is the mass density. All lengths are normalized to a generalized minor radius $a$; the resistivity to $\eta_0$ (its value at the magnetic axis); the time to the Alfv\' en time $\tau_A = R_0 (\mu_0 \rho_m)^{1/2} / B_0$; the magnetic field to $B_0$ (the averaged value at the magnetic axis); and the pressure to its equilibrium value at the magnetic axis $p_0$. The Lundquist number $S$ is the ratio of the resistive time $\tau_R = a^2 \mu_0 / \eta_0$ to the Alfv\' en time. The $\beta$-value at the magnetic axis is $ \beta_{0} = 2 \mu_{0} p_{0} / B_{0}^2 $.

Each equation has a perpendicular dissipation term, with the characteristic coefficients $D$ (the collisional cross-field transport), and $\mu$ (the collisional viscosity for the perpendicular flow). A source term $Q$ is added to equation (\ref{peq}) to balance the energy losses.

Equilibrium flux coordinates $(\rho, \theta, \zeta)$ are used. Here, $\rho$ is a generalized radial coordinate proportional to the square root of the toroidal flux function, and normalized to one at the edge. The flux coordinates used in the code are those described by Boozer \cite{25}, and $\sqrt g$ is the Jacobian of the coordinates transformation.  All functions have equilibrium and perturbation components like $ A = A_{eq} + \tilde{A} $. The operator $ \nabla_{||} $ denotes  derivation in the  direction parallel to the magnetic field, and is defined as

\begin{equation*}
\nabla_{||} = \frac{\partial}{\partial\zeta} + \rlap{-} \iota\frac{\partial}{\partial\theta} - \frac{1}{\rho}\frac{\partial\tilde{\psi}}{\partial\theta}\frac{\partial}{\partial\rho} + \frac{\partial\tilde{\psi}}{\partial\rho}\frac{1}{\rho}\frac{\partial}{\partial\theta},
\end{equation*}
where $\rlap{-} \iota$ is the rotational transform.

The FAR3D code uses finite differences in the radial direction and Fourier expansions in the two angular variables. The poloidal flux $\psi$ and the pressure $p$ are expressed as cosine series,
\begin{equation*}
\tilde f (\rho, \theta, \zeta, t)=  \sum_{n,m} \tilde{f}_{n,m}(\rho, t)\cos(m\theta + n\zeta),
\end{equation*}
and the stream function $\Phi$ is expressed as a sine series,
\begin{equation*}
\tilde \Phi (\rho, \theta, \zeta, t)=  \sum_{n,m} \tilde{\Phi}_{n,m}(\rho, t)\sin(m\theta + n\zeta).
\end{equation*}
The numerical scheme is semi-implicit in the linear terms. The nonlinear version uses a two semi-steps method to ensure $(\Delta t)^2$ accuracy.

\subsection{Equilibrium properties}

\begin{figure}
\includegraphics{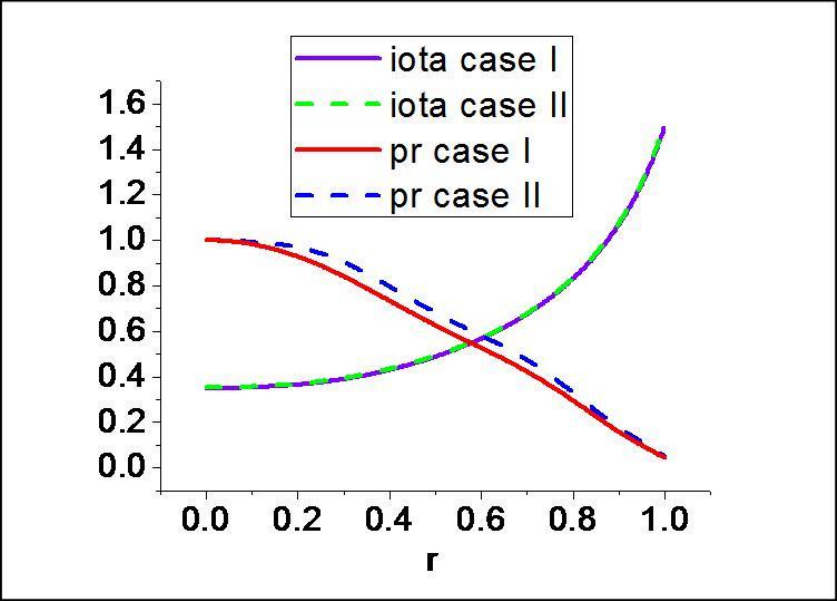}
\caption{\label{FIG:1} Pressure profile and rotational transform for cases I and II (before and after the sawtooth like event, respectively).} 
\end{figure}

Two different equilibria have been used for the simulations. Both were calculated by using the free-boundary version of the VMEC code \cite{19}. The electron density and temperature profiles where reconstructed by Thomson scattering data and electron cyclotron emission. This is a high density plasma produced by sequentially injected hydrogen pellets, and strongly heated by 3 NBI
after the last pellet injection. The first equilibrium (case I) is calculated from experimental data measured before a sawtooth like event, and the second equilibrium (case II) is calculated from data after the same event \cite{9}. The vacuum magnetic axis is inward-shifted ($R_{{\rm{axis}}} = 3.6$ m), the magnetic field at the magnetic axis is $2.75$ T, the inverse aspect ratio $\varepsilon$ is $0.16$, and $\beta_0$ is $1.48 \%$ for case I, and $1.25 \%$ for case II.
The pressure profile in case I is more peaked than in case II (see figure~\ref{FIG:1}), and the rotational transform profile is similar for both equilibria.

\subsection{Calculation parameters}

\begin{figure}[h]
\includegraphics{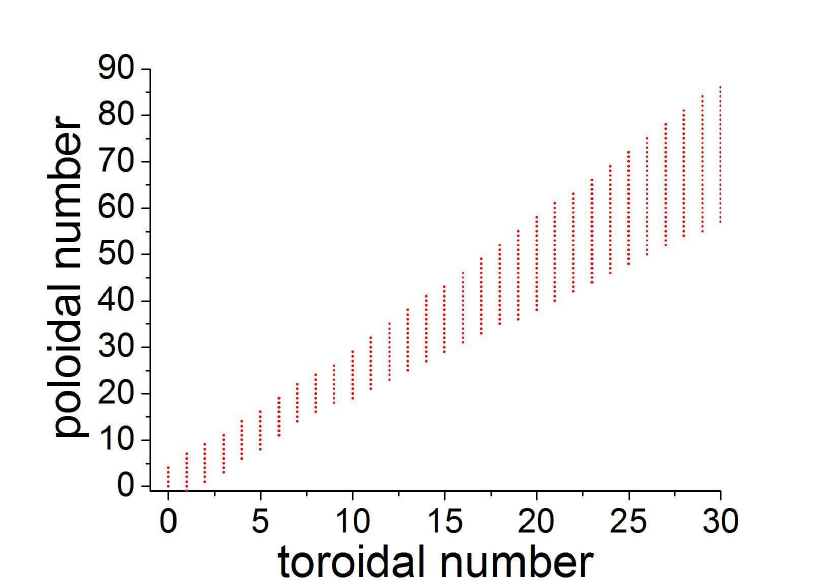}
\caption{\label{FIG:2} Fourier modes included in the simulation.} 
\end{figure}

The calculations have been done with a uniform radial grid of 500 points. Up to 515 Fourier components have been included in the calculations (see figure~\ref{FIG:2}). The maximum $n$ value is 30. The FAR3D code allows for linear helical couplings. However, for the equilibria considered here, the linear results obtained including these helical couplings do not change substantially with respect to those obtained including only toroidal couplings.
For this reason, in these nonlinear calculations we only include equilibrium components with $n = 0$ and $0 \le m \le 5$. The Lundquist number is $S=10^5$, and the coefficients of the dissipative terms are independent of $\rho$ with values $\mu=7.5 \times 10^{-6}$ and $D=1.25 \times 10^{-5}$. They are normalized to $a^2/\tau_A$. The resistivity is also assumed constant ($\eta =1$).

The Lundquist number is $2$-$3$ orders of magnitude lower than the experimental value in the plasma core. For $S = 10^{5}$ the plasma has a larger resistivity than in the experiment. A previous research \cite{10} has shown that  sawtooth like events are observed  for $S < 10^{8}$. The $S$ value is smaller than the value in the experiment for computational reasons, and the consequence is that sawtooth events in the simulation will be stronger than those observed in the experiments \cite{23, 26}.

\begin{figure}[h]
\includegraphics{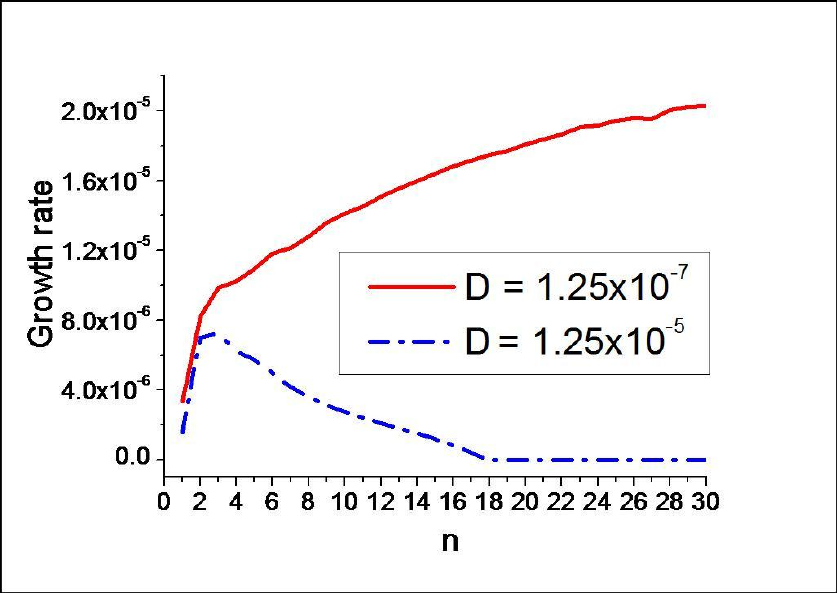}
\caption{\label{FIG:3} Linear growth rate vs $n$ for two different values of the diffusion coefficient $D$.} 
\end{figure}

Figure~\ref{FIG:3} shows the effect of diffusion on the linear growth rate of each toroidal mode family. Modes with toroidal mode number $ n > 3 $ suffer a strong stabilization but the linear growth rate of low $n$ modes ($n = 1,2,3$) is only weakly affected by the dissipation. With $D=1.25 \times 10^{-5}$ we have a significant range of stable modes ($18 \le n \le 30$).

In order to reach a smooth saturation, we increase gradually $\beta$ in the simulation. We start with a $\beta$-value half the final value and we increment the value in five steps. The increment is one tenth of the final value.  From now on, we will denote the first period of the evolution (half the final $\beta$) as $A$, the second (3/5 of the final $\beta$) as $B$, and so on, until reaching the final $\beta$-value in period $F$. The final $\beta$-values are the experimental ones, that is, $\beta_0=1.48 \%$ for case I, and $\beta_0=1.25 \%$ for case II.

The source term $Q$ added to equation (\ref{peq}) is a Gaussian centered near the magnetic axis. This energy input is dynamically fitted, in such a way that the value of the volume integral of the pressure is kept almost constant during the evolution.

\subsection{Diagnostics}

We use several diagnostics to characterize the evolution. The averaged pressure profile, $\left\langle p \right\rangle = p_{\rm{eq}}(\rho) + \tilde{p}_{00} (\rho)$, where the angular brackets indicate average over a flux surface, and $\tilde{p}_{00}$ is the $(n=0, m=0)$ Fourier component of the pressure perturbation. The unstable modes drive deformations in the pressure profile in the vicinity of the rational surfaces, hence the averaged pressure $\left\langle p \right\rangle$ flattens in the more unstable regions. The instantaneous rotational transform profile is given by the expression
\begin{equation}
\label{iota}
\rlap{-} \iota (\rho)+ \tilde{\rlap{-} \iota}(\rho) = \rlap{-} \iota+ \frac{1}{\rho}\frac{\partial\tilde{\psi}}{\partial\rho}
\end{equation}
This instantaneous rotational transform profile gives us information about the resonant modes in the plasma and the (instantaneous) position of their rational surfaces. This diagnostic will be crucial to understand fluctuations close to the magnetic axis.

Two-dimensional contour plots are very useful to visualize regions of large gradients and strong perturbations. We will use three versions of contour plots. The first one is the contour plot of the pressure. Expressed in terms of the Fourier expansion, $p = p_{eq}(\rho) + \sum_{n,m} \tilde{p}_{n,m}(\rho)\cos (m\theta + n\zeta)$. Sometimes it will be more interesting to use the pressure perturbation, $ \tilde{p} = \sum_{n,m} \tilde{p}_{n,m}(\rho)\cos (m\theta + n\zeta) $, or the perturbation of the pressure with respect to the (flux surface) averaged pressure, $ p - \left\langle p \right\rangle = \sum_{(n,m) \not= (0,0)} \tilde{p}_{n,m}(\rho)\cos (m\theta + n\zeta) $.

The last set of diagnostics is related to the magnetic field structure. By integrating the magnetic field line equations, we can obtain Poincar\'e plots at a given toroidal angle. This Poincar\'e plots are very useful to visualize the topology of the (instantaneous) magnetic field. By including in the Fourier expansion of the magnetic flux $\psi$ only the components belonging to a given helicity $n/m$, we get the magnetic island corresponding to that helicity. When the magnetic islands overlap,  some stochastic regions appear where the magnetic field lines will cover some volume of the torus \cite{20}. These stochastic magnetic field line structures are visualized when all the Fourier modes are included in the expansion of $\psi$, and these stochastic regions are associated to different rational surfaces very close between them but not always overlapped.

\section{Simulation results \label{sec:simulation}}

\begin{figure}[h]
\includegraphics{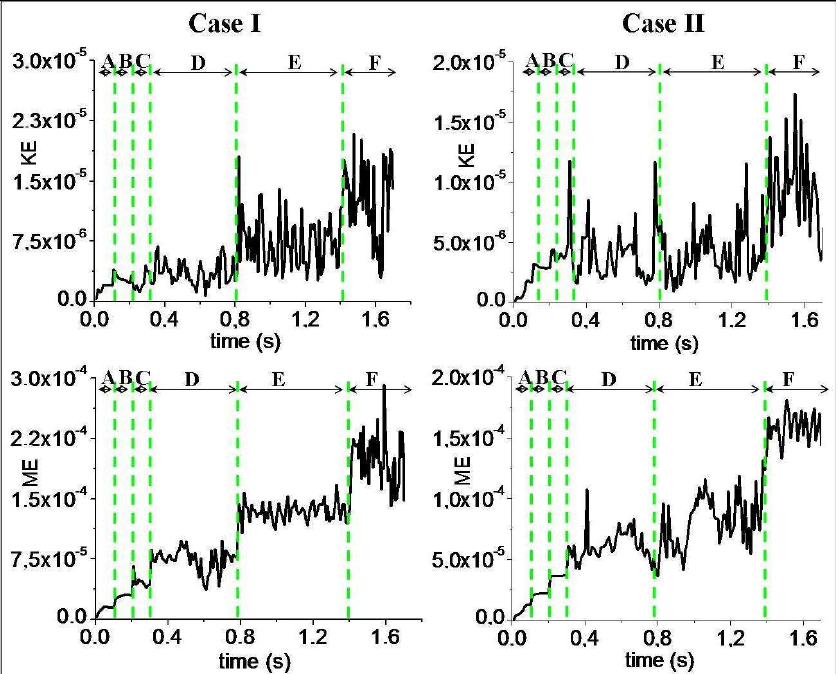}
\caption{\label{FIG:4} Kinetic (up) and magnetic (down) energy evolution for cases I (left) and II (right). Green lines indicate the transition between different periods of the simulation.}
\end{figure}

For each $\beta$-value, fluctuations nonlinearly evolve to a saturated state. The energy at saturation increases as the $\beta$-value raises (see figure~\ref{FIG:4}). In periods A to C the saturation level is almost constant with small oscillations. For periods D to F, there are strong oscillations in the steady state. There are some overshoots when $\beta$ is changed (transitions from one period to the next) but  the evolution is smooth most of the time.  In all the calculations, we assume that the resistive time $\tau_R$ is 1 second.

\begin{figure}[h]
\includegraphics{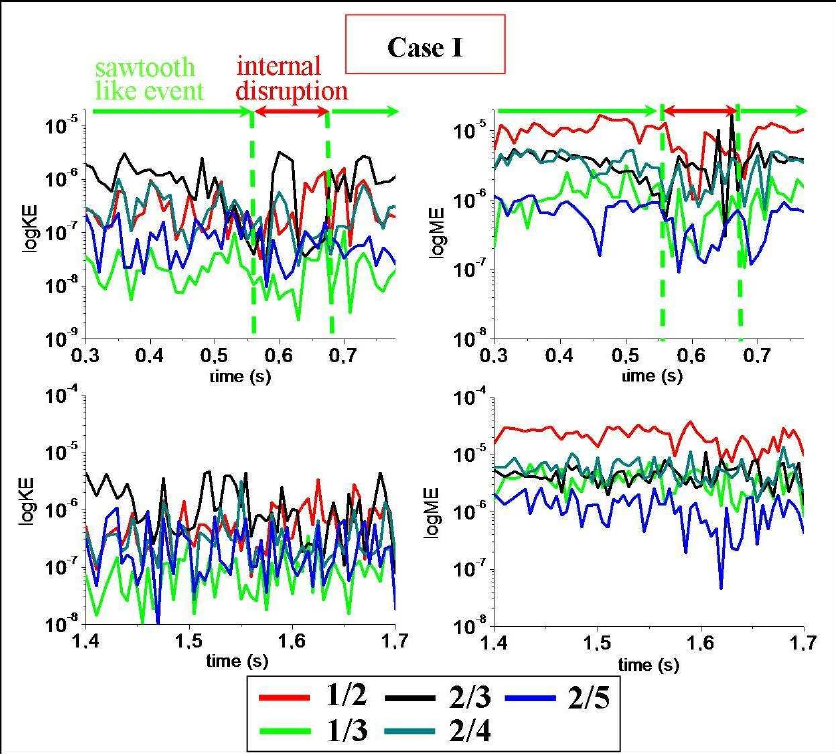}
\caption{\label{FIG:5} Kinetic (left) and magnetic (right) energy evolution of the dominant modes for case I during  periods D (up) and F (down). Green arrows denote the time intervals with non resonant sawtooth like events and red arrows the time interval with $1/2$ internal disruptions.}
\end{figure}

The energy evolution of the dominant modes during period D is shown in figure~\ref{FIG:5} (up) for case I. We distinguish two different types of events during the evolution: non resonant sawtooth like events and internal disruptions. Sawtooth like events are related with fast changes in the energy of dominant modes  $n/m = 1/2 $, $ 2/3 $ and magnetic energy local maxima of mode $ 1/3 $.  We call them ``non resonant'' because the 1/3 rational surface is not in the plasma. Internal disruptions occur at the 1/2 rational surface, and  they are preceded by a fast decrease of the energy of the dominant modes  before a sharply increase of the energy of modes $ 2/3 $ and $2/5$, with a local minimum of the energy of mode $1/3$.

The evolution of the energy of the dominant modes  in period F (see figure~\ref{FIG:5} down) is similar to periods D and E but the oscillations are stronger and faster because the system has more energy to drive instabilities. In this period there are no internal disruptions but a new class of sawtooth like event is observed when the mode $1/3$  is located inside the plasma core. We call such event resonant sawtooth like event.

\begin{figure}[h]
\includegraphics{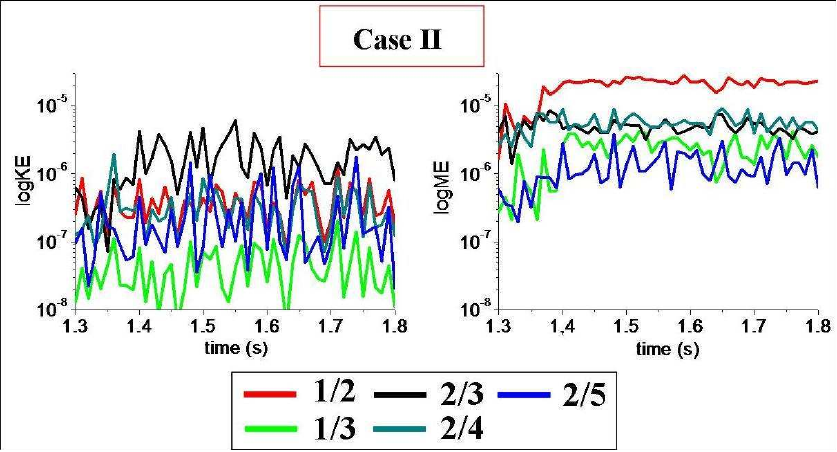}
\caption{\label{FIG:6} Kinetic (left) and magnetic (right) energy evolution of the dominant modes for case II during  period F.}
\end{figure}

The energy evolution during period F of case II is shown in figure~\ref{FIG:6}. The oscillations and the instabilities are weaker than in the case I. In this period only non resonant sawtooth like events are driven.

In summary, non resonant sawtooth like events and internal disruptions are observed for $ \beta_{0} \geq 1 \% $ for cases I and II. Resonant sawtooth like events are only observed for case I and $ \beta_{0} = 1.48 \%$. Internal disruptions were not observed for the latter. The properties of each type of event are described in the next subsections.

\subsection{Non resonant sawtooth like event, case I}

\begin{figure}[h]
\includegraphics{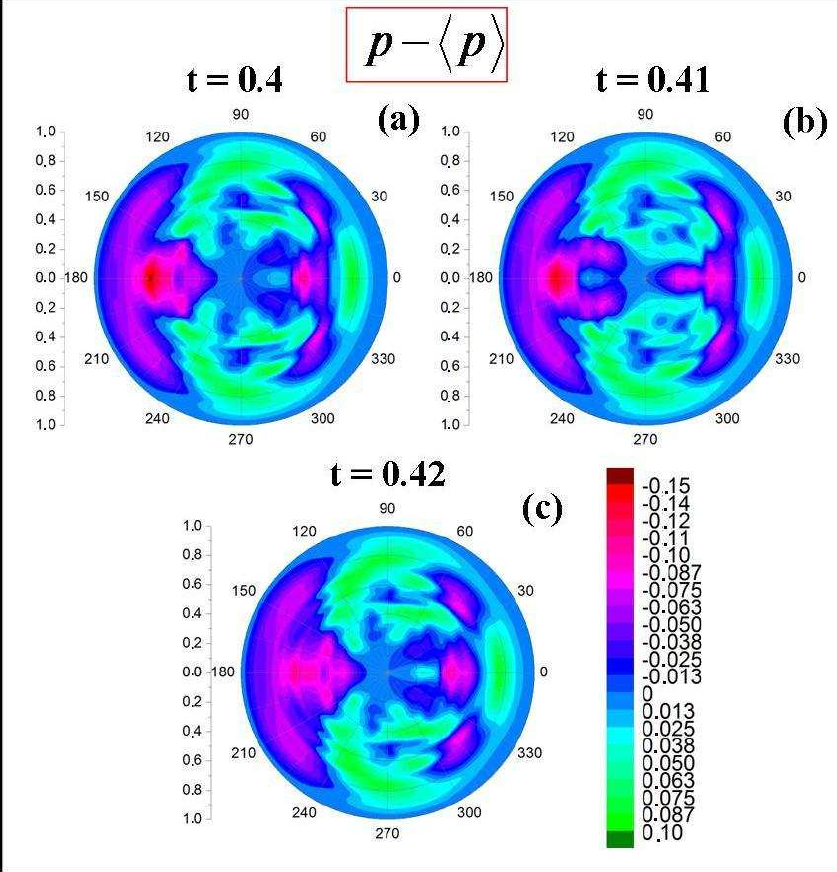}
\caption{\label{FIG:7} Instantaneous contours of pressure for a non resonant sawtooth like event of case I. (a) to (c) Perturbation with respect to the averaged pressure.}
\end{figure}

\begin{figure}[h]
\includegraphics{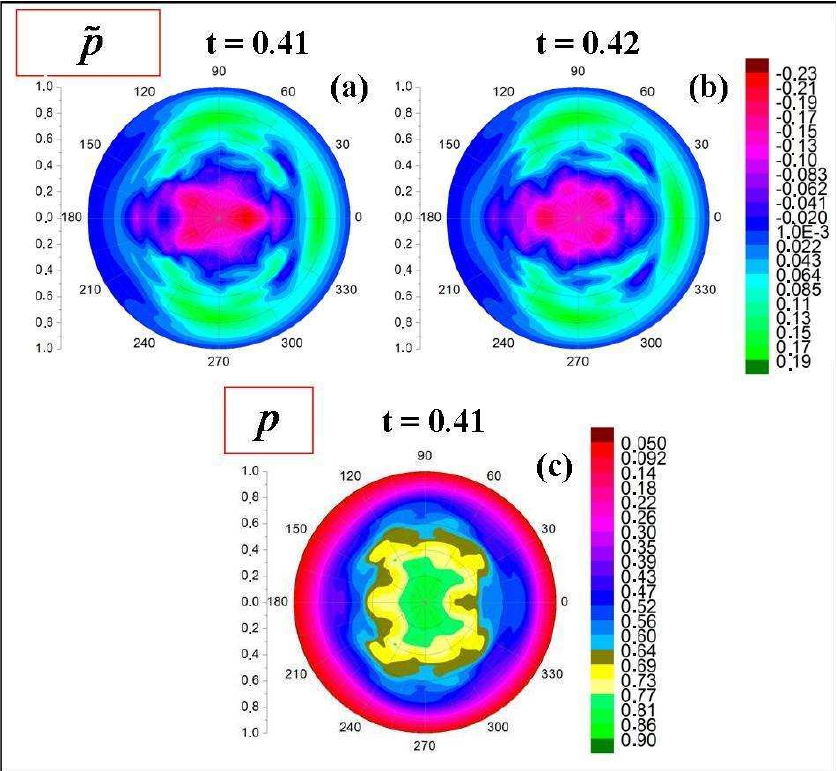}
\caption{\label{FIG:8} Instantaneous contours of pressure for a non resonant sawtooth like event of case I. (a) and (b) Perturbation with respect to the equilibrium. (c) Total pressure.}
\end{figure}

Figures~\ref{FIG:7} and~\ref{FIG:8}  show the evolution of pressure iso-contours on the poloidal plane $\zeta = 0$ to illustrate a non resonant sawtooth like event of case I. Since we use Boozer coordinates to display these contours, the equilibrium flux surfaces are circles.

At $t=0.4$ s, Fig.~\ref{FIG:7}(a), the instability is located in the middle region of the plasma.  Between $t=0.4$ and $t=0.41$ s, the instability propagates to the inner region. At $t = 0.41$ s, Fig.~\ref{FIG:7}(b), the middle and inner part of the plasma are linked by an $m=3$ structure due to the non resonant effect of mode $1/3$, and the magnetic energy of the 1/3 component reaches a local maximum.   At $t=0.42$ s, Fig.~\ref{FIG:7}(c), the dominant instability is again located in the middle region of the plasma. At both $t=0.4$ and $t=0.42$ s, the deformation in the middle region of the plasma is an $m=5$ structure due to the  $2/5$ component. The perturbation with respect to the equilibrium shows a pattern of three blobs in the inner plasma region at $t = 0.41$ s, Fig.~\ref{FIG:8}(a), and a pattern of five blobs at $t = 0.42$ s, Fig.~\ref{FIG:8}(b). Therefore the $1/3$ component dominates in the inner region at $t = 0.41$ s, but the $2/5$ component dominates at $t = 0.50
 $ s. The iso-contours of the total pressure at $t=0.42$ s are shown in Fig.~\ref{FIG:8}(c). The contours are strongly deformed with respect to the equilibrium, but there is not a clear pattern corresponding to a given component.

\begin{figure}[h]
\includegraphics{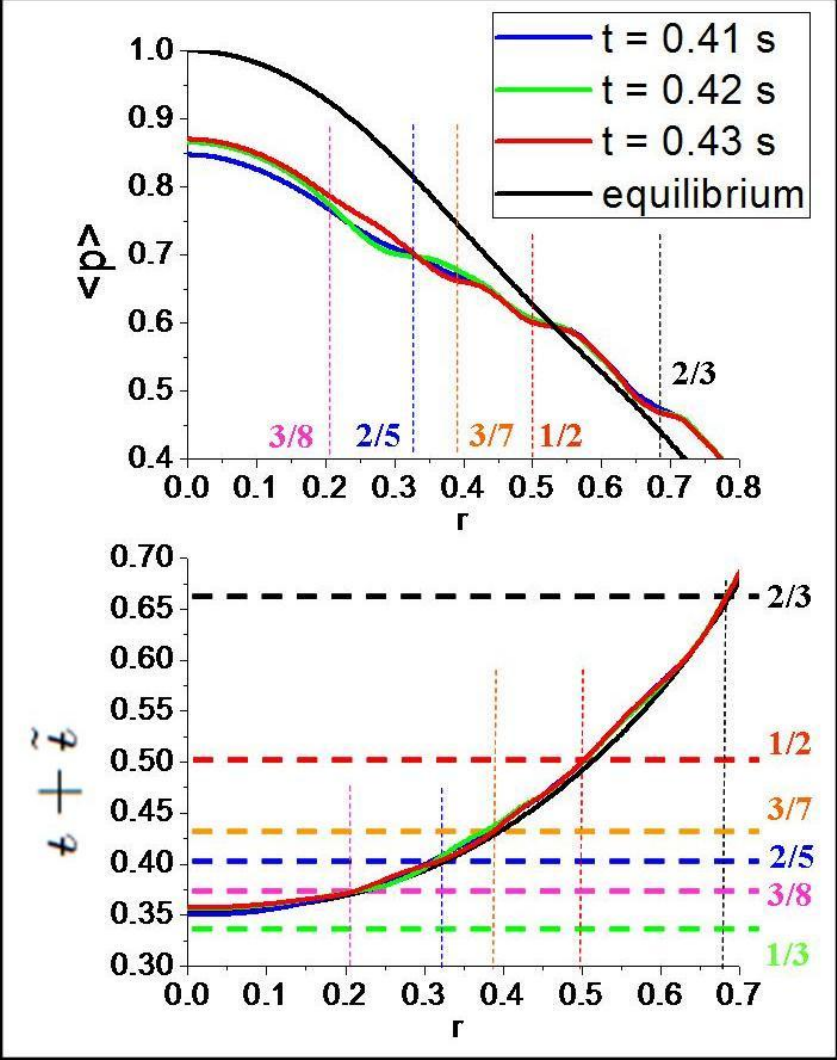}
\caption{\label{FIG:9} Evolution of the averaged pressure (top) and rotational transform (bottom) profiles for a non resonant sawtooth like event, case I.}
\end{figure}

The averaged pressure and the instantaneous rotational transform profiles during the non resonant sawtooth like event are shown in figure~\ref{FIG:9}.  At $t = 0.41$ s, the pressure profile flattening in the inner region corresponds to the $2/5$ mode, in the middle region to  the $1/2$ and $3/7$ modes, and in the periphery to the $2/3$ mode. At $t = 0.42$ s,  the flattening at the $3/7$ rational surface decreases but the flattening at the  $2/5$ rational surface slightly increases. It seems from these observations that the deformation of the pressure profile in the region $0 < \rho < 0.2$  at $t = 0.41$ s is not driven by the $2/5$ rational surface, because its maximum deformation takes place at $t = 0.42$ s, and its effect is located in the region $0.25 < \rho < 0.4$. This leaves us  with the non resonant $1/3$ component as the main candidate to explain the deformation in the region $0 < \rho < 0.2$.

\begin{figure}[h]
\includegraphics{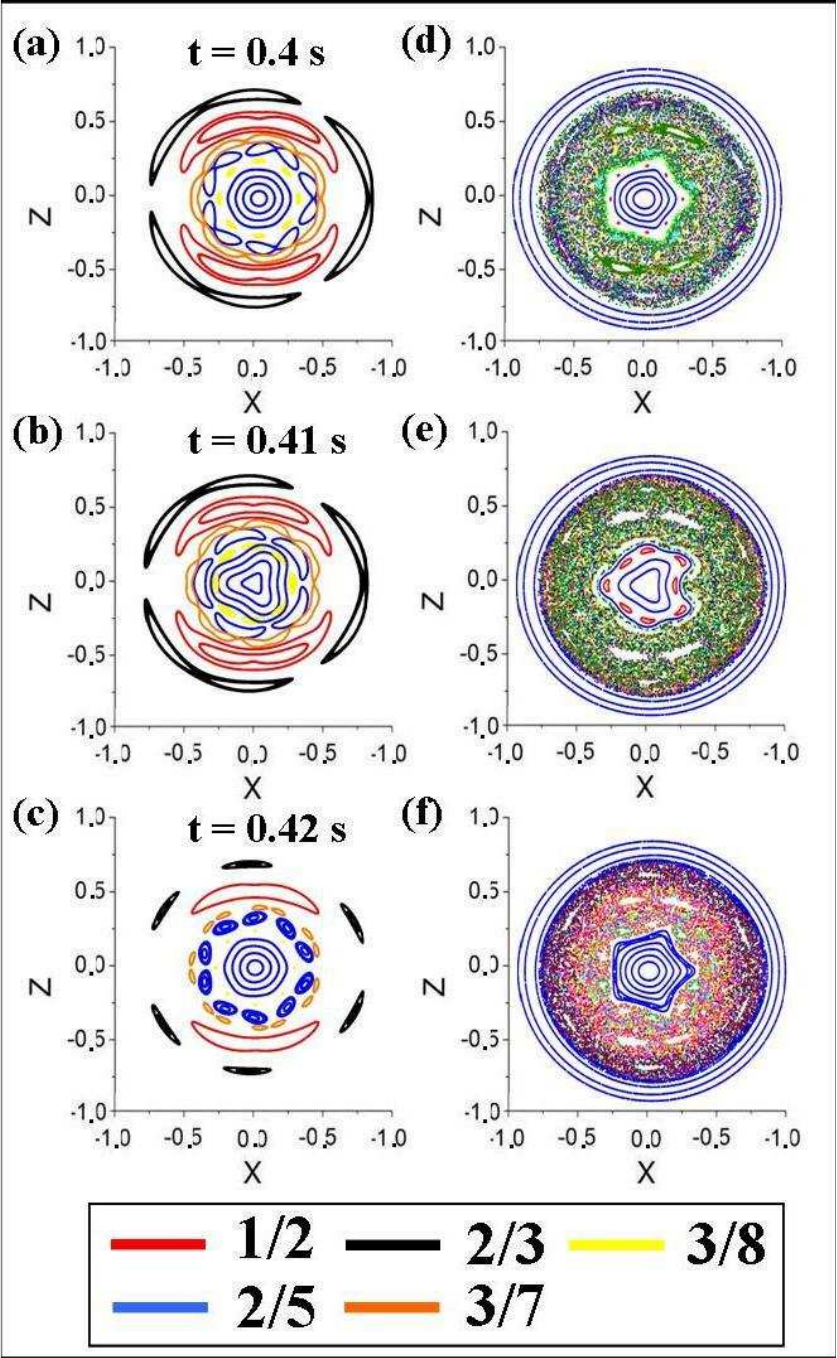}
\caption{\label{FIG:10} Field line plots on the poloidal plane $\zeta=0$ for a non  resonant sawtooth like event, case I. (a) to (c) Including only modes belonging to each main helicity in the Fourier series. (d) to (f) Including all the modes in the Fourier series.}
\end{figure}

The magnetic field structure during the non resonant sawtooth like event is shown in figure~\ref{FIG:10}. For each time, we obtain the Poincar\'e plots in two ways. First we include only the modes belonging to a given helicity $n/m$ in the Fourier series of the perturbed poloidal flux. We repeat the calculation for the helicity of the main components. This gives us plots which represent the structure of the magnetic islands associated to each helicity, like in Figs.~\ref{FIG:10}(a) to (c). When all the modes are included in the calculation, we get the total magnetic field line plots, Figs.~\ref{FIG:10}(d) to (f).

At $t = 0.4$ and $t = 0.41$ s a stochastic field region appears between the inner and outer region due to the overlap of the islands with helicities $1/2$, $2/5$, $3/7$ and $3/8$. At $t = 0.41$ s the $1/3$ mode non resonant effect drives a strong deformation of the magnetic surfaces near the magnetic axis, and the stochastic region reaches the periphery. At $t = 0.42$ the magnetic island sizes decrease, and the deformation is reduced near the magnetic axis, decreasing the stochastic region in the periphery. The $2/5$ island does not reach its maximum size at $t = 0.41$ s when the $1/3$ mode deformation in the plasma core is maximum. Therefore, during the non resonant sawtooth events the main contribution to the instability is possibly due to the $1/3$ mode.

In summary, when the maximum energy of the $ 1/3 $  mode is reached, and the rotational transform at the magnetic axis is close to $1/3$, a non resonant sawtooth like event can be driven between the middle and inner plasma regions. After non resonant sawtooth like events the equilibria does not suffer a main modification and these oscillations can be driven consecutively. The system efficiency to confine the plasma decreases only slightly, because the stochastic region is located only between the middle and inner plasma, but it does not reach the inner part or the plasma periphery.

It is worth noting that the Lundquist number used in the simulations, $S=10^5$, is two orders of magnitude smaller than the experimental value. As a consequence, the magnetic island sizes are going to be much larger than the experimental ones \cite{23, 26}.

\subsection{Non resonant sawtooth like event, case II}

\begin{figure}[h]
\includegraphics{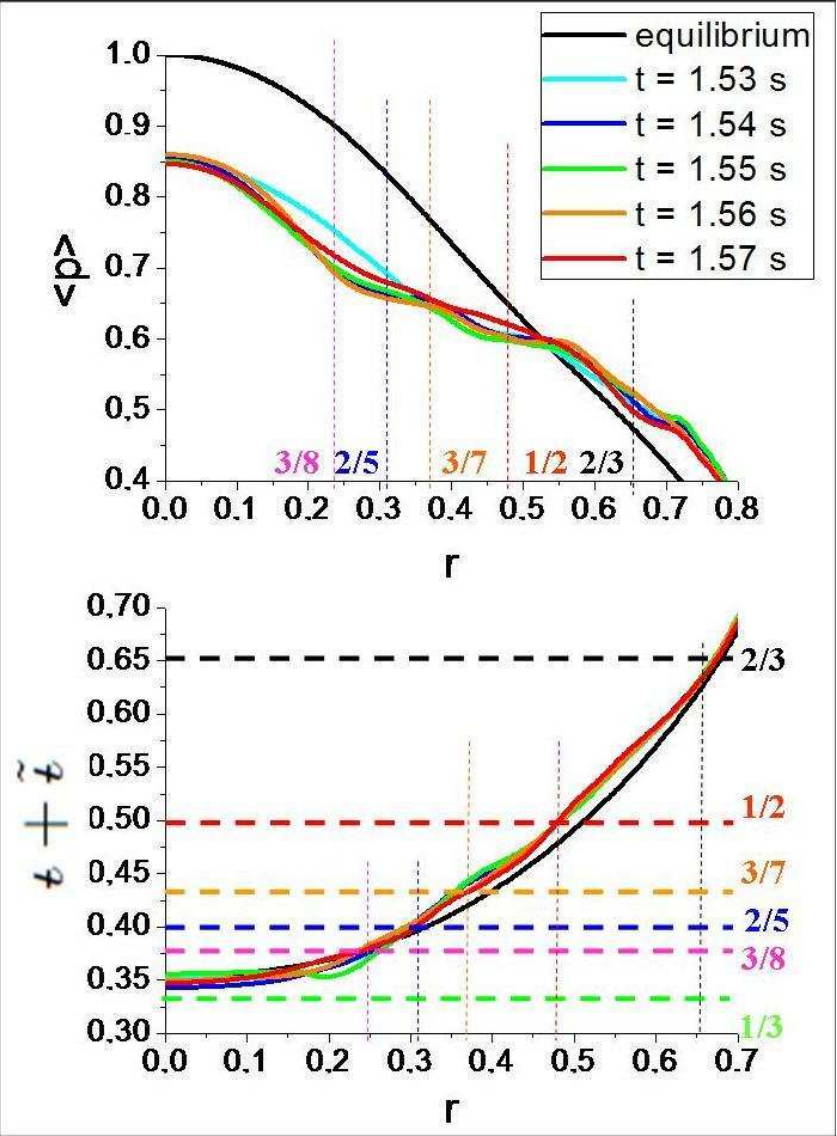}
\caption{\label{FIG:11} Evolution of the averaged pressure (top) and rotational transform (bottom) profiles for a non resonant sawtooth like event, case II.}
\end{figure}

The equilibrium pressure profile near the magnetic axis in case II  is flatter than in case I. Therefore case II is more stable than case I,  sawtooth events are weaker and their evolution is slower. In case II, period F (experimental $\beta$ value), only non resonant sawtooth like events are driven. Figure~\ref{FIG:11} shows the averaged pressure and the instantaneous rotational transform profiles during a non resonant sawtooth like event in case II, period F.

\begin{figure}[h]
\includegraphics{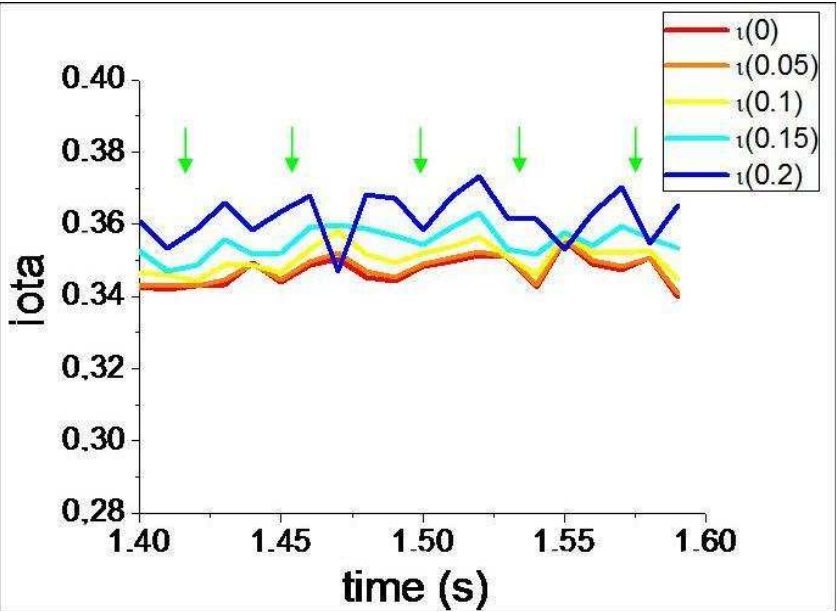}
\caption{\label{FIG:12} Evolution of the $\rlap{-} \iota$ value at $ \rho = 0, 0.05, 0.1, 0.15, 0.2 $. Green arrows show the non resonant events.}
\end{figure}

The sawtooth event starts at $ t = 1.53 $ s. The averaged pressure profile is flattened in the inner plasma region between $ t = 1.54 $ and $ t = 1.56 $ s. The event ends at $ t = 1.57 $ s when the pressure profile is less flattened in the inner region. The pressure profile flattening is similar to  case I but the deformation does not reach the plasma region close to the magnetic axis ($ \rho < 0.25$).

The number of events driven during the simulation period F in  case II is lower than in  case I. The evolution of the value of the rotational transform at $ \rho = 0, 0.05, 0.1, 0.15$ and $0.2 $ is shown in figure~\ref{FIG:12}. The beginning of each non resonant sawtooth event is indicated by a green arrow.

\begin{figure}[h]
\includegraphics{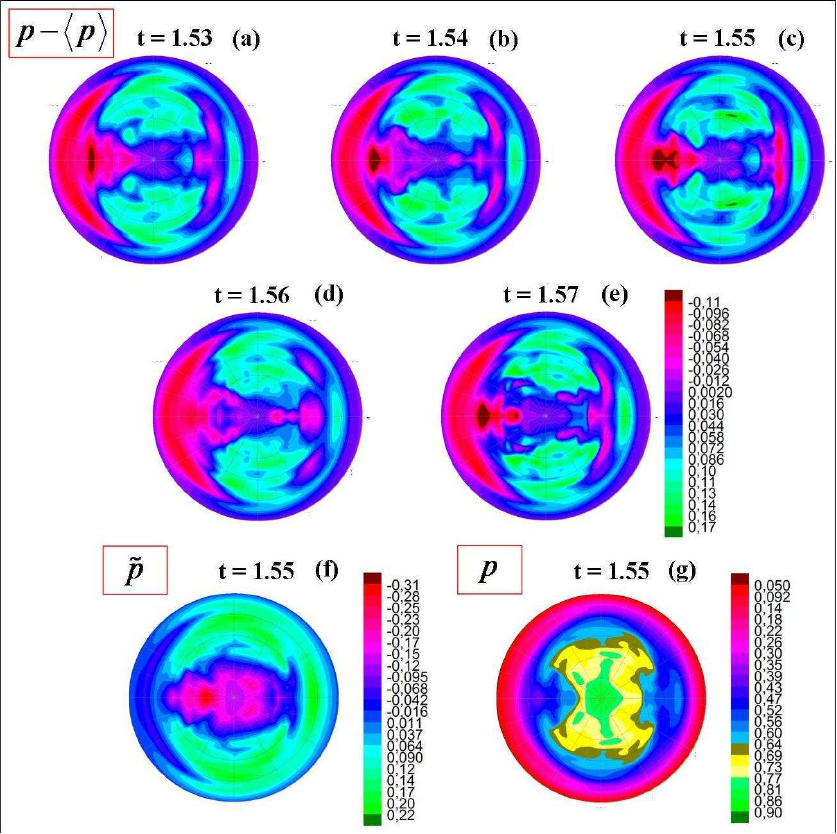}
\caption{\label{FIG:13} Instantaneous contours of pressure for a non resonant sawtooth like event of case II. Perturbation with respect to the averaged pressure (a) to (e), perturbation with respect to the equilibrium (f) and total pressure (g).}
\end{figure}

The pressure iso-contours on the poloidal plane $\zeta = 0$, figure~\ref{FIG:13} (a) to (e), show similar patterns than the case I non resonant events but the evolution is slower. At $ t = 1.54 $ and $ 1.56 $ s the instability propagates between the middle plasma and the inner plasma region like a $m=3$ structure. At $ t = 1.55 $ s the pressure gradient deformation is maximum in the inner plasma but the instability is closer to the middle plasma region than in the case I. At $ t = 1.57 $ s, the $m=3$ structure disappears and a $m=5$ structure dominates in the inner plasma.

The pressure perturbation with respect to the equilibrium shows a pattern of three blobs in the inner plasma region at $t = 1.55$ s, Fig.~\ref{FIG:13} (f), but the structures are not as closer to the magnetic axis as in case I . The iso-contours of the total pressure at $t=1.55$ s, Fig.~\ref{FIG:13} (g), show a large flux surface deformation in the middle plasma like in the case I but it is weaker near the magnetic axis.

\begin{figure}[h]
\includegraphics{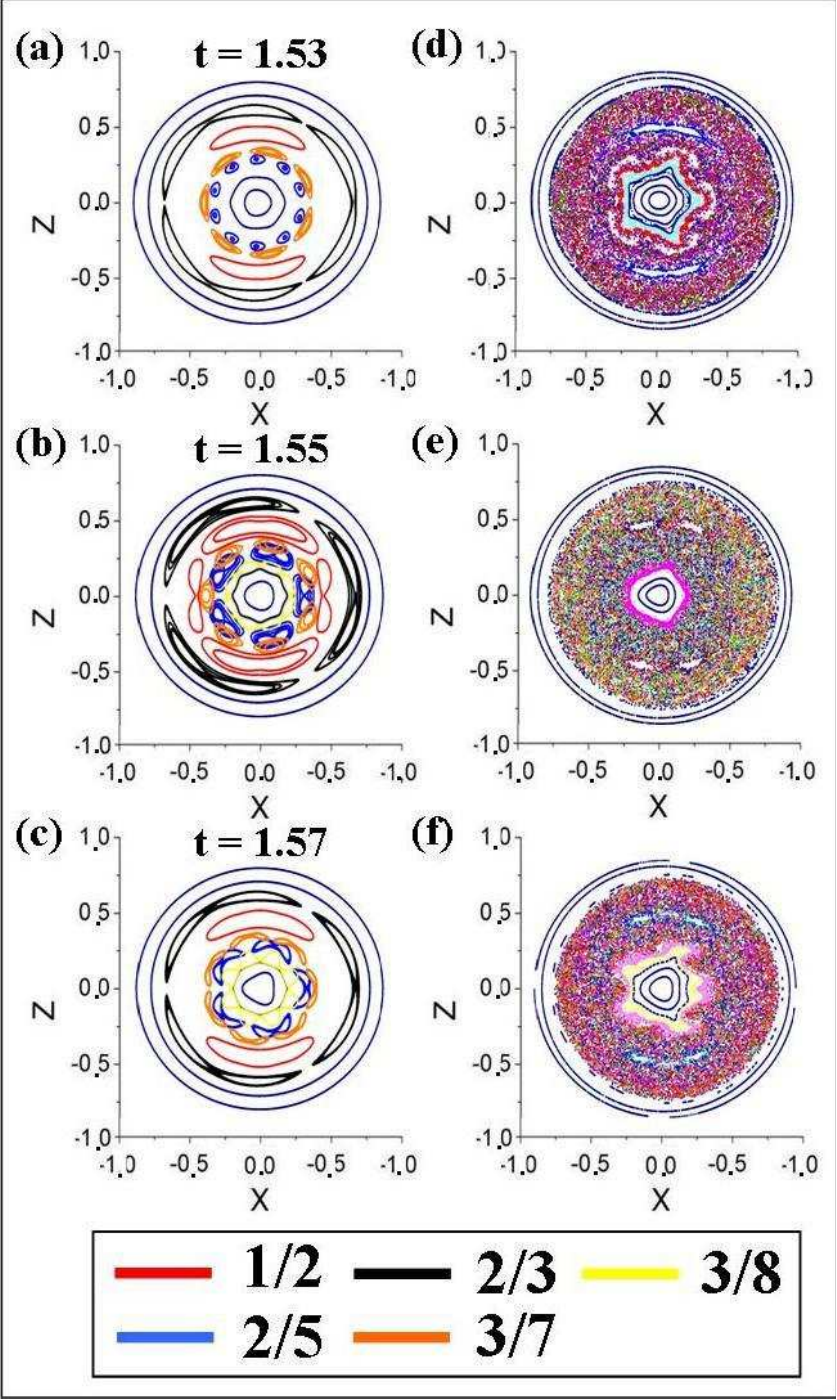}
\caption{\label{FIG:14} Field line plots on the poloidal plane $\zeta=0$ for a non  resonant sawtooth like event, case II. (a) to (c) Including only modes belonging to each main helicity in the Fourier series. (d) to (f) Including all the modes in the Fourier series.}
\end{figure}

The magnetic field structure, Fig.~\ref{FIG:14}, is again similar to the case I patterns. At $ t = 1.53 $ s the dominant magnetic islands overlapping is small and the stochastic region in the middle of the plasma does not reach the inner region. At $ t = 1.55 $ s the magnetic island size increases and the stochastic region reaches the inner region; several magnetic surfaces break down in the inner plasma region and are perturbed in the region close to the magnetic axis. In case I, the stochastic region is closer to the magnetic axis, where the distortion of the magnetic surfaces is larger. At $ t = 1.57 $ s, the stochastic region and the magnetic islands size decrease but the magnetic surfaces are recovered slower than in case I.

\subsection{Internal disruption}

\begin{figure}[h]
\includegraphics{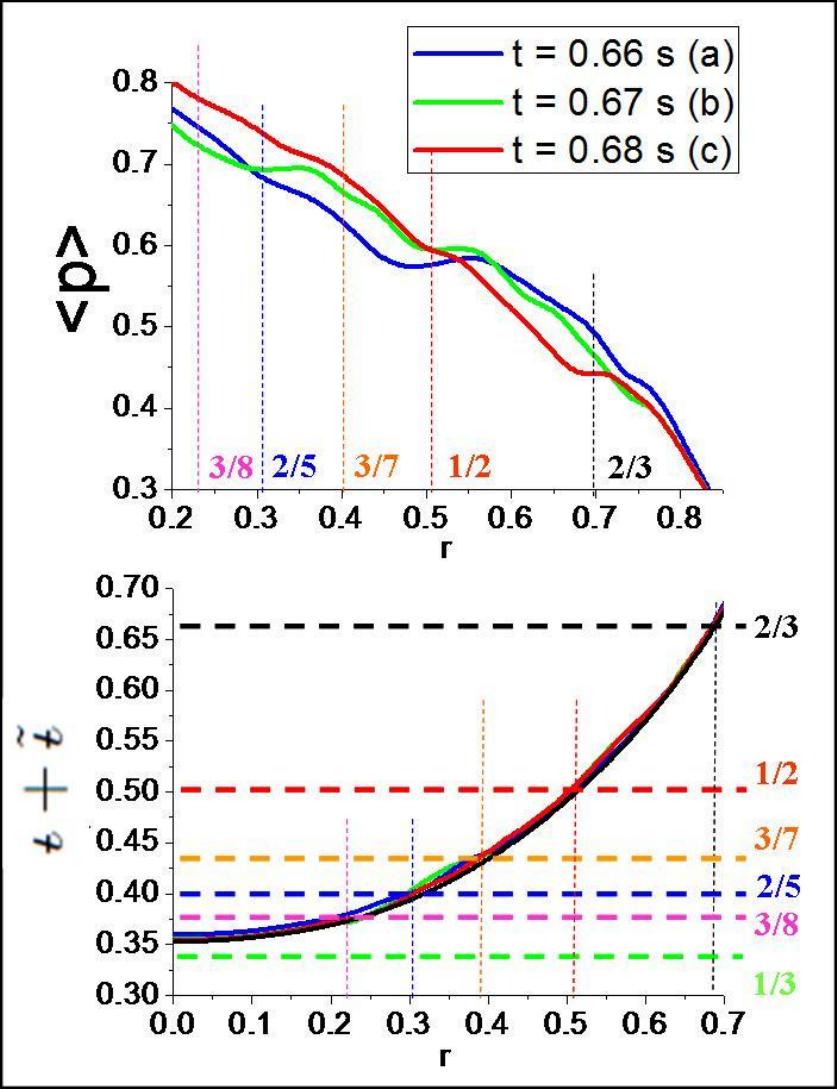}
\caption{\label{FIG:15} Evolution of the averaged pressure (top) and rotational transform (bottom) profiles for an internal disruption event.}
\end{figure}

Next, we examine the so-called internal disruption event that is preceded by a fast decrease of the $1/2 $ mode energy (see figure~\ref{FIG:5}). From the averaged pressure profile, figure~\ref{FIG:15}, we see that there is a strong deformation in $\rho = 0.5$  at $t = 0.66$ s. This deformation is located at the position of  the $\rlap{-} \iota=1/2$ rational surface,  where the profile shows an inversion. At $t = 0.67$ s the flattening of the profile in the middle region decreases and appears a new profile flattening at $\rho = 0.3$ driven by the rational surface $2/5$. At $t = 0.68$ s the profile flattening disappears in the inner region and decreases in the middle region of the plasma, while in the periphery the deformation driven by the rational surface $2/3$ is large. From this, we can conclude that the internal disruption starts with a strong deformation due to the $1/2$ mode. Later on, this perturbation propagates in radius in both directions.

\begin{figure}[h]
\includegraphics{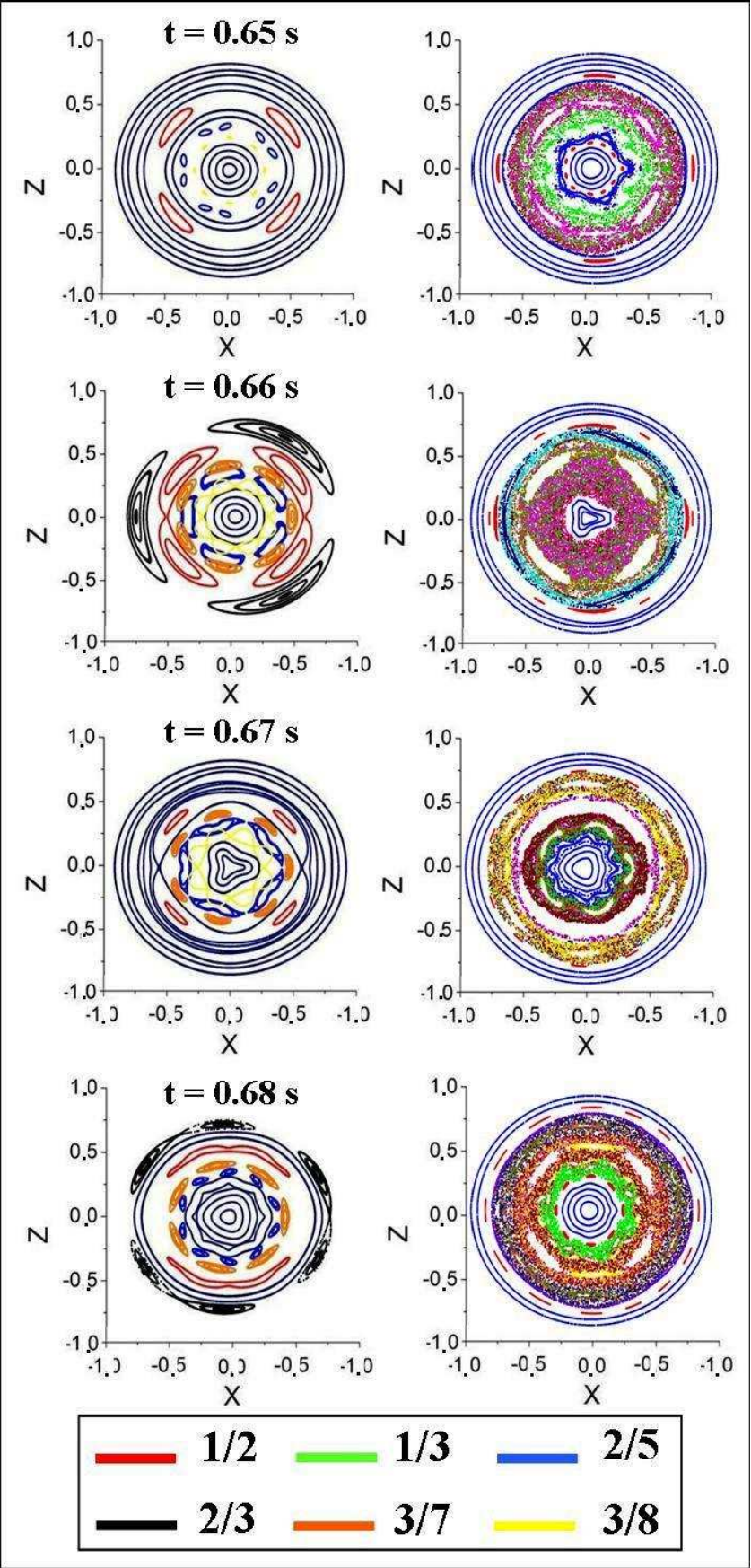}
\caption{\label{FIG:16} Field line plots on the poloidal plane $\zeta=0$ for an internal disruption event. Including only modes belonging to each main helicity in the Fourier series (up). Including all the modes in the Fourier series (down).}
\end{figure}

The magnetic field structure during the internal disruption event is shown in figure~\ref{FIG:16}. At $t = 0.65$ s the island sizes are small and there is no strong overlapping. There are several distinct regions where the magnetic surfaces are well defined between the inner region and the periphery. At $t = 0.66$ s there are broad magnetic islands of different helicicities that overlap between the inner and the periphery plasma region. The $1/2$ and $2/3$  islands clearly dominate.  At $t = 0.67$ s a sort of magnetic reconnection takes place in the  middle region of the plasma. There is still large overlapping of the islands and the magnetic surface deformation close to the magnetic axis is strong. At $t = 0.68$ s the magnetic reconnection reaches the plasma inner region, and  the deformation close to the magnetic axis disappears, recovering some good plasma confinement regions like at $ t = 0.65 $ s.

In summary, the $1/2$ helicity is the main driver of  internal disruptions. The dominant deformation in the averaged pressure is that associated with the $1/2$ component. The equilibrium in an internal disruption suffers an important reorganization, because the stochastic region expands along the plasma. Unlike sawtooth like events, we have not observed consecutive internal disruptions. The instability in an internal disruption starts in the middle region of the plasma, and expands to the inner region and the periphery of the plasma.

\subsection{Resonant sawtooth like event}

\begin{figure}[h]
\includegraphics{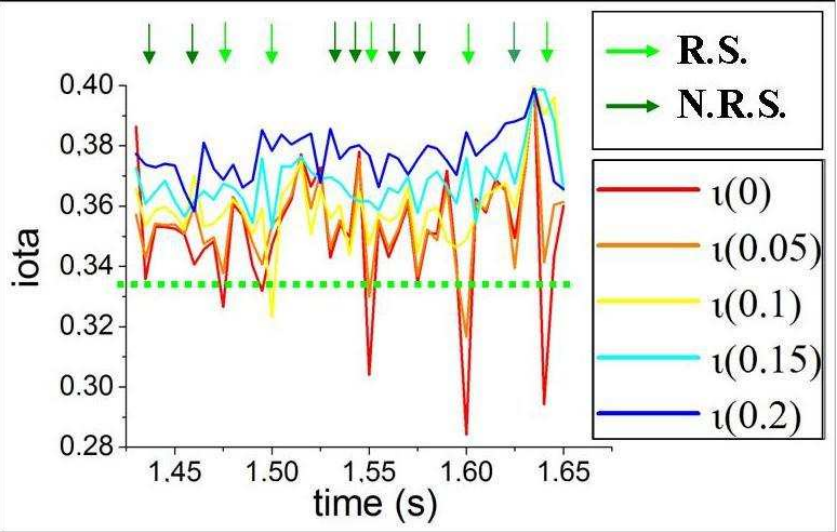}
\caption{\label{FIG:17} Evolution of the $\rlap{-} \iota$ value at $ \rho = 0, 0.05, 0.1, 0.15, 0.2 $. Green dotted line shows the $ \rlap{-} \iota = 1/3 $. The light (dark) green arrows show the resonant (non resonant) events.}
\end{figure}

For case I and $ \beta_{0} = 1.48 \%$, the $1/3$ rational surface is located inside the plasma during some phases of the sawtooth like events. This is illustrated in figure~\ref{FIG:17}, where the evolution of the rotational transform at the magnetic axis and at  $ \rho = 0.05, 0.1, 0.15, 0.2 $ is shown. The time interval between peaks shows the periodicity of the sawtooth event.

\begin{figure}[h]
\includegraphics{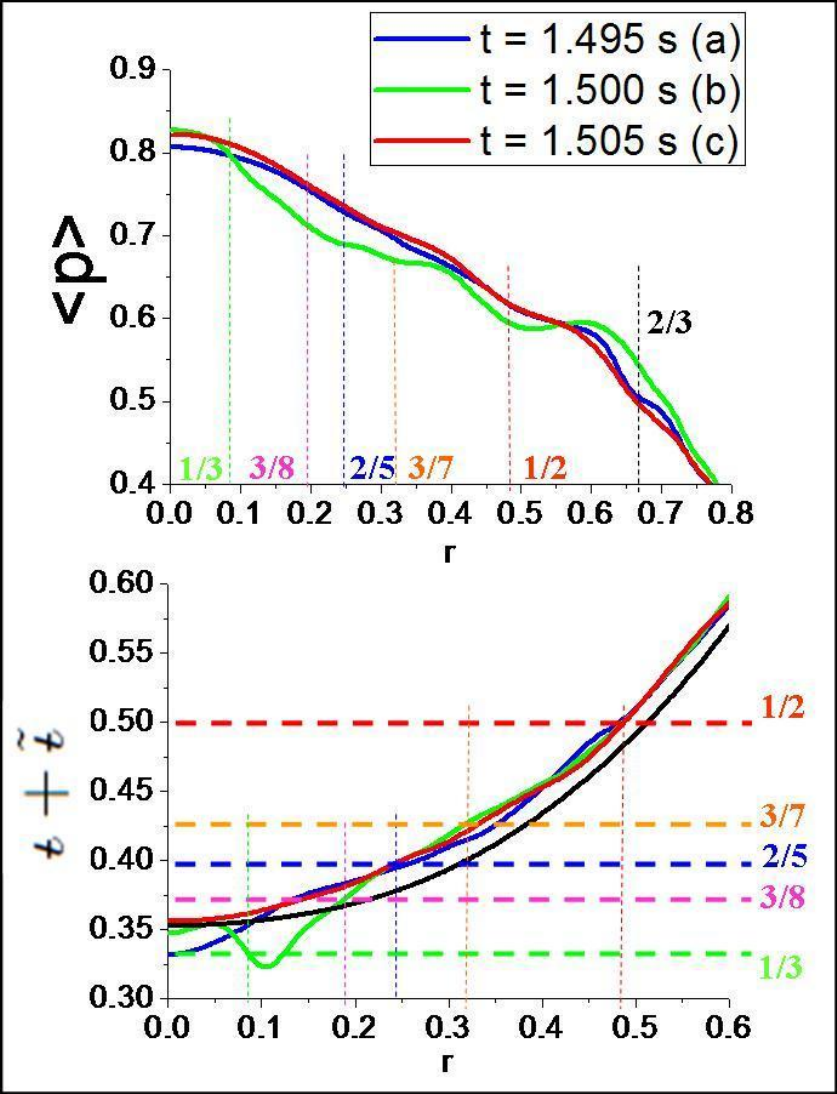}
\caption{\label{FIG:18} Evolution of the averaged pressure and rotational transform profiles for a resonant sawtooth like event.}
\end{figure}

Figure~\ref{FIG:18} shows that the instantaneous rotational transform profile crosses the $1/3$ value at $t=1.5$ s, and the averaged pressure profile is strongly modified.

\begin{figure}[h]
\includegraphics{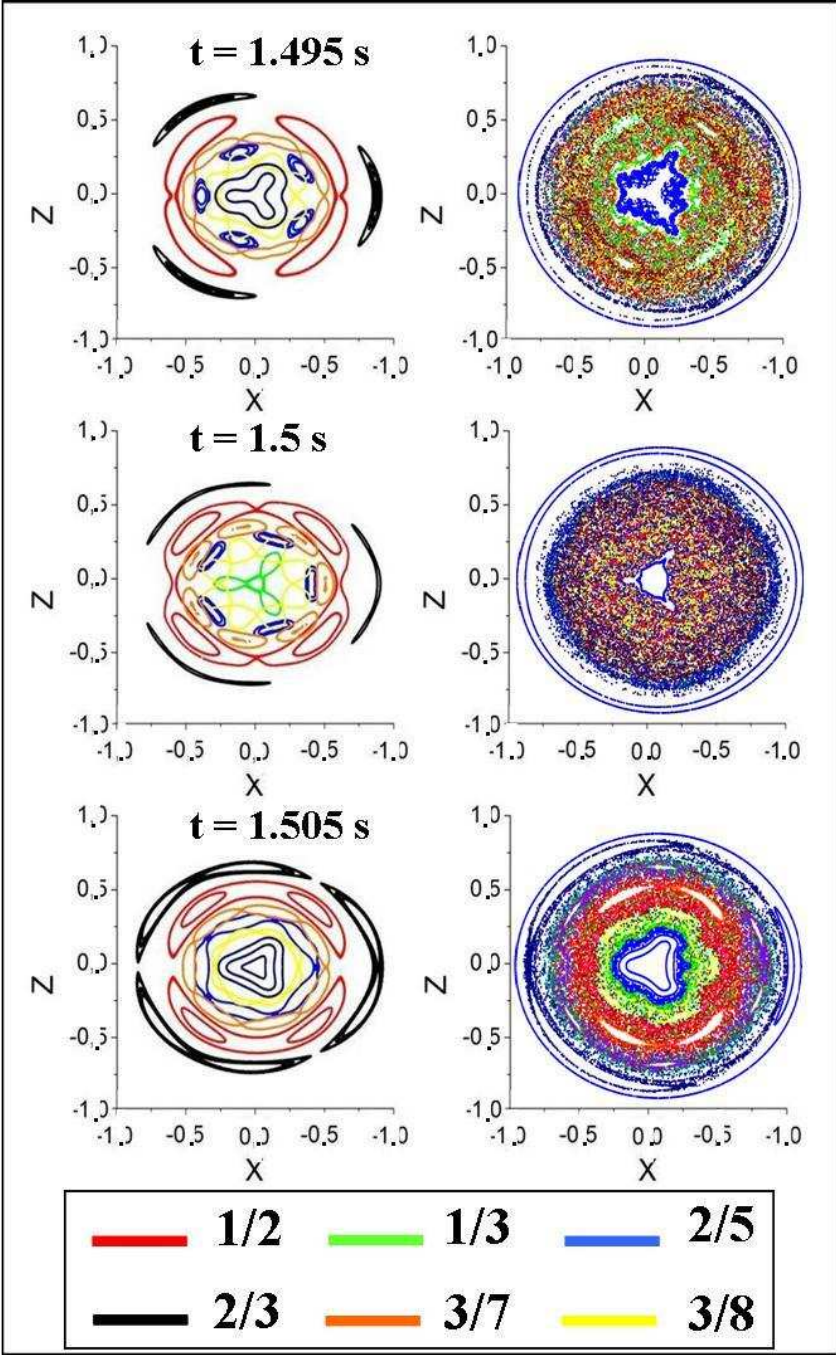}
\caption{\label{FIG:19} Field line plots on the poloidal plane $\zeta=0$ for a  resonant sawtooth like event. Including only modes belonging to each main helicity in the Fourier series (up). Including all the modes in the Fourier series (down).}
\end{figure}

The magnetic field structure during the resonant sawtooth like event is shown in figure~\ref{FIG:19}. At $ t = 1.495 $ and $1.505$ s the magnetic islands $2/5$, $3/8$, $3/7$ and $1/2$ slightly overlap between the periphery and the inner region, with several regions where the magnetic surfaces are well defined. At $t = 1.5$ s the  $1/3$  resonant mode drives a strong magnetic deformation close to the magnetic axis, and magnetic islands associated to the $1/3$ helicity appear. The overlap of these islands with the other islands present in the plasma broadens the stochastic region until reaching most of the plasma volume.

In summary, in case I with $ \beta_{0} = 1.48 $ $ \% $ the sawtooth like events are more intense because the rational surface $1/3$ is inside the plasma. During period F ($ \beta_{0} = 1.48  \% $) internal disruptions are not driven because the deformation near the magnetic axis due to the $1/3$ component prevents the deformation due to the $1/2$ mode reaches the plasma inner region. The equilibrium does not suffer a large reorganization after a resonant sawtooth like event, and consecutive events can be driven.  The $1/3$ magnetic islands close to the magnetic axis expand the stochastic region until reaching most of the plasma volume.

\section{Conclusions and discussion \label{sec:conclusions}}

The results of the simulations show that internal disruptions and non resonant sawtooth like events are driven for $\beta_{0} \geq 1\%$, but resonant sawtooth like events are only driven for case I and $\beta_{0} = 1.48\%$. The main driver of sawtooth like events is the $n/m = 1/3$ mode. Non resonant events correspond to instances in which the $\rlap{-} \iota=1/3$ rational surface is not present in the plasma, and resonant events to instances in which the $\rlap{-} \iota=1/3$ rational surface is in the plasma, close to the magnetic axis. The resonant sawtooth drives a larger instability than the non resonant, and both events are stronger as $\beta$ increases. The sawtooth like events are less dangerous than the internal disruptions driven by the $1/2$ mode, because the equilibrium suffers a major reset after an internal disruption. Internal disruptions are not driven in the simulations with $\beta_{0} = 1.48 \%$, because the $1/3$ mode causes a strong deformation, and the deformation driven by the  $1/2$ mode does not reach the inner region of the plasma. The resonant and non resonant sawtooth like events frequency increases with $\beta$, and for $\beta_{0} = 1.48 \%$ these events can remove the system energy before an internal disruption is driven.

The Lundquist number in the simulations is $2$-$3$ orders of magnitude lower than the experimental value. Because the simulation is more resistive than the experiment, the disruption and sawtooth like events driven in the simulation are stronger than the sawtooth like activity observed in the experiments. The relationship between internal disruptions and $1/2$ sawtooth like activity will be the target of a future research.

During LHD operation, non resonant sawtooth like events are driven and they cannot be avoided, because the $1/3$ non resonant effect will always influence the stability in the inner region of the plasma and it can be only reduced by keeping the rational transform profile far away from $\rlap{-} \iota=1/3$. The resonant events can be prevented if the rotational transform profile does not fall bellow $1/3$.

\begin{acknowledgments}
This research was sponsored in part by the DGICYT (Direcci\'on General de Investigaciones Cient{\'\i}ficas y Tecnol\'ogicas) of Spain under project No. ENE2009-12213-C03-03.
\end{acknowledgments}


\begin{thebibliography}{10}

\bibitem{1} B. A. Carreras, H. R. Hicks, J. A. Holmes, V. E. Lynch, L. Garcia, J. H. Harris, T. C. Hender, and B. F. Masden, {\it Phys. Fluids} {\bf 26}, 3569 (1983).
\bibitem{2} Y. Nakamura, M. Wakatani, J-N. Leboeuf, B. A. Carreras, N. Domingnez, J. A. Holmes, V. E. Lynch, S. L. Painter and L. Garcia, {\it Fusion Technol.} {\bf 19}, 217 (1991).
\bibitem{3} MOTOJIMA Osamu, YAMADA Hiroshi, ASHIKAWA Naoko, EMOTO Masahiko, FUNABA Hisamichi, GOTO Motoshi, IDA Katsumi, IDEI Hiroshi, IKEDA Katsunori, INAGAKI Shigeru, INOUE Noriyuki, ISOBE Mitsutaka, KANEKO Osamu, KAWAHATA Kazuo, KHLOPENKOV Konstantin, KOBUCHI Takashi, KOMORI Akio, KOSTRIOUKOV Artem, KUBO Shin, KUMAZAWA Ryuhei, LIANG Yunfeng, MASUZAKI Suguru, MINAMI Takashi, MIYAZAWA Junichi, MORISAKI Tomohiro, MORITA Shigeru, MURAKAMI Sadayoshi, MUTO Sadatsugu, MUTO Takashi, NAGAYAMA Yoshio, NAKAMURA Yukio, NAKANISHI Hideya, NARIHARA Kazumichi, NARUSHIMA Yoshiro, NISHIMURA Kiyohiko, NODA Nobuaki, NOTAKE Takashi, OHDACHI Satoshi, OHYABU Nobuyoshi, OKA Yoshihide, OSAKABE Masaki, OZAKI Tetsuo, PETERSON Byron, SAGARA Akio, SAITO Kenji, SAKAKIBARA Satoru, SAKAMOTO Ryuichi, SASAO Mamiko, SATO Kuninori, SATO Motoyasu, SEKI Tetsuo, SHIMOZUMA Takashi, SHOJI Mamoru, SUZUKI Hajime, TAKEIRI Yasuhiko, TANAKA Kenji, TAMURA Naoki, TOI Kazuo, TOKUZAWA Tokihiko, TORII Yuki, TSUMORI Katsuyo
 shi, WATANABE Kiyomasa, WATANABE Tsuguhiro, YAMADA Ichihiro, YAMOMOTO Satoshi, YOKOYAMA Masayuki, YOSHIMURA Yasuo, WATARI Tetsuo, XU Yuhoug, CHIKARAISHI Hirotaka, HAMAGUCHI Shinji, HISHINUMA Yoshimitsu, IMAGAWA Shinsaku, IWAMOTO Akifumi, MAEKAWA Ryuji, MITO Toshiyuki, NISHIMURA Arata, TAMURA Hitoshi, YAMADA Shuichi, YANAGI Nagato, TAKAHATA Kazuya, ITOH Kimitaka, MATSUOKA Keisuke, OHKUBO Kunizo, SATOW Takashi, SUDO Shigeru, UDA Tatsuhiko and YAMAZAKI Kozo, {\it J. Plasma Fusion Res. SERIES} {\bf 5}  22 (2002).
\bibitem{4} K. Ichiguchi, Y. Nakamura, M. Wakatani, N. Yanagi and S. Morimoto, {\it Nucl. Fusion} {\bf 29}, 2093 (1989).
\bibitem{5} Ichiguchi K., {\it  Proc. 1999 Intl. Stellarator Workshop Madison}, CD-ROM, file P2-4  (1999).
\bibitem{7} K.Y. Watanabe, S. Sakakibara, Y. Narushima, H. Funaba, K. Narihara, K. Tanaka, T. Yamaguchi, K. Toi1, S. Ohdachi, O. Kaneko, H. Yamada, Y. Suzuki, W.A. Cooper, S. Murakami, N. Nakajima, I. Yamada, K. Kawahata, T. Tokuzawa, A. Komori and LHD experimental group, {\it Nucl. Fusion} {\bf 45}, 1247 (2005).
\bibitem{8} K. Ichiguchi, M. Wakatani, T. Unemura, T. Tatsuno and B.A. Carreras, {\it  Nucl. Fusion} {\bf 41},  181 (2001).
\bibitem{6} K. Ichiguchi, N. Nakajima, M. Wakatani, B.A. Carreras and V.E. Lynch, {\it Nucl. Fusion} {\bf 43}, 1101 (2003).
\bibitem{9} S. Ohdachi, K. Toi, G. Fuchs, S. Sakakibara, K. Y. Watanabe, Y. Narushima, K. Narihara, K. Tanaka, T. Tokuzawa, S. Inagaki, Y. Nagayama, F. Watanabe, K. Kawahata, K. Komori and the LHD Exprimental Group, {\it 21 IAEA Fusion Energy Conference, 16 - 21 October Chengdu, China}, EX/P8-15 (2006).
\bibitem{10} S. Ohdachi, K. Y. Watanabe, S. Sakakibara, H. Yamada, Y. Narushima, H. Funaba, Y. Suzuki, K. Toi, I. Yamada, T. Minami, K. Narihara, K. Tanaka, T. Tokuzawa, K. Kawahata, A. Komori and the LHD Experimental Group, {\it 22nd IAEA Fusion Energy Conference}, {\it http://www-pub.iaea.org/MTCD/Meetings/fec2008pp.asp}, EX/8-1Rb (2008).
\bibitem{11} S Ohdachi, S Yamamoto, K Toi, S Sakakibara, K Y Watanabe, Y Narushima, H Yamada, I Yamada, K Narihara, K Tanaka, K Tokuzawa, K Kawahata, and LHD Experimental Group, {\it Proceedings of 13th stellarator workshop 2002, Canberra}, PIIA.9 (2002).
\bibitem{12} Y. Nagayama, K. Kawahata, S. Inagaki, B. J. Peterson, S. Sakakibara, K. Tanaka, T. Tokuzawa, K. Y. Watanabe, N. Ashikawa, H. Chikaraishi, M. Emoto, H. Funaba, M. Goto, Y. Hamada, K. Ichiguchi, K. Ida, H. Idei, T. Ido, K. Ikeda, S. Imagawa, A. Isayama, M. Isobe, A. Iwamoto, O. Kaneko, S. Kitagawa, A. Komori, S. Kubo, R. Kumazawa, S. Masuzaki, K. Matsuoka, T. Mito, J. Miyazawa, T. Morisaki, S. Morita, O. Motojima, S. Murakami, T. Mutoh, S. Muto, N. Nakajima, Y. Nakamura, H. Nakanishi, K. Narihara, Y. Narushima, A. Nishimura, K. Nishimura, A. Nishizawa, N. Noda, S. Ohdachi, K. Ohkubo, N. Ohyabu, Y. Oka, M. Osakabe, T. Ozaki, A. Sagara, K. Saito, R. Sakamoto, M. Sasao, K. Sato, T. Seki, T. Shimozuma, M. Shoji, H. Suzuki, S. Sudo, K. Takahata, Y. Takeiri, K. Toi, K. Tsumori, H. Yamada, I. Yamada, K. Yamazaki, N. Yanagi, M. Yokoyama, Y. Yoshimura, Y. Yoshinuma, T. Watari, and LHD Groupl, {\it Phys. Rev. Lett.} {\bf 90}, 205001 (2003).
\bibitem{13} J. H. Harris, O. Motojima, H. Kaneko, S. Besshou, H. Zushi, M. Wakatani, F. Sano, S. Sudo, A. Sasaki, K. Kondo, M. Sato, T. Mutoh, T. Mizuuchi, M. Iima, T. Obiki, A. Iiyoshi, and K. Uo, {\it Phys. Rev. Lett.} {\bf 53}, 2242 (1984).
\bibitem{14} H. Zushi, Y. Suzuki, M. Hosotsubo, Y. Nakamura, M. Wakatani, F. Sano, K. Kondo, T. Mizuuchi, M. Nakasuga, S. Besshou, H. Okada, K. Nagasaki, C. Christou, Y. Kurimoto, H. Funaba, T. Hamada, T. Kinosita, T. Obiki, S, Kado, K. Muraoka, S. Sudo, K. Ida, B. J. Peterson, V. Yu. Sergeev, K. V. Khlopenkov, V. V. Chechkin, V. S. Voitsenja, \textit{Proceedings of the 16th International Conference on Fusion Energy, IAEA, Montreal, 1996}, Vol.2, pp 143-150 (1997).
\bibitem{15} B. A. Carreras, V. E. Lynch, H. Zushi, K. Ichiguchi, and M. Wakatani, {\it Phys. Plasmas} {\bf 5}, 3700 (1998).
\bibitem{16} S.Ohdachi, {\it Proc. 24th EPS Conference on Contr. Fusion and Plasma Phys, Berchtesgaden}, 21A, 817 (1997).
\bibitem{17} S. Takagi, K. Toi, M. Takechi, S. Murakami, K. Tanaka, S. Nishimura, M. Isobe, K. Matsuoka, T. Minami, S. Okamura, M. Osakabe, C. Takahashi, and Y. Yoshimura, {\it Phys. Plasmas} {\bf 11}, 1537 (2004).
\bibitem{18} L. Garcia, \textit{Proceedings of the 25th EPS International Conference, Prague, 1998}, VOL. 22A, Part II, p. 1757.
\bibitem{21} L.A Charlton, J.A Holmes, H.R Hicks, V.E Lynch and B.A Carreras, {\it Journal of Comp. Physics} {\bf 63}, 107 (1986).
\bibitem{22} L.A Charlton, J.A Holmes, V.E Lynch, B.A Carreras and T.C Hender, {\it Journal of Comp. Physics} {\bf 86}, 270 (1990).
\bibitem{19} S. P. Hirshman and J. C. Whitson {\it Phys. Fluids} {\bf 26}, 3553 (1983).
\bibitem{24} L. Garcia, B. A. Carreras, N. Dominguez, J. N. Leboeuf, and V. E. Lynch, {\it Phys. Fluids B} {\bf 2}, 2162 (1990).
\bibitem{25} A.H. Boozer, {\it Phys. Fluids} {\bf 25}, 520 (1982).
\bibitem{23} S. SAKAKIBARA, K. Y. WATANABE, H. YAMADA, Y. NARUSHIMA, T. YAMAGUCHI, K. TOI, S. OHDACHI, A. WELLER, K. TANAKA, K. NARIHARA, K. IDA, T. TOKUZAWA, K. KAWAHATA, A. KOMORI, and LHD Experimental Group, {\it Fusion Sci. Technol.} {\bf 50}, 177 (2006).
\bibitem{26} B. A. Carreras, L. Garcia, and P. H. Diamond, {\it Phys. Fluids } {\bf 30}, 1388 (1987).
\bibitem{20} B.V. Chirikov, {\it Phys. Rev. Lett. } {\bf 5}, 263 (1979).


\end{thebibliography}
\end{document}